\documentclass[12pt]{article}

 \usepackage{amssymb}
 \usepackage{amsmath}
 \usepackage{amsfonts}
 \usepackage{color}
 \newcommand {\be}{\begin{equation}}
 \newcommand {\ee}{\end{equation}}
 \newcommand {\bea}{\begin{array}}
 
 \newcommand {\eea}{\end{array}}

\begin{document}

 %\begin{titlepage}
  \thispagestyle{empty}

  \vspace{2cm}

  \begin{center}
    \font\titlerm=cmr10 scaled\magstep4
    \font\titlei=cmmi10 scaled\magstep4
    \font\titleis=cmmi7 scaled\magstep4
  {\bf Symplectic Quantization of \\Massive Bosonic string in  background B-field}

    \vspace{1.5cm}
    \noindent{{%\large
    A. Shirzad\footnote{shirzad@ipm.ir}
        A. Bakhshi\footnote{abakhshi@ph.iut.ac.ir}
      Y. Koohsarian%\footnote{koohsarian_ramian@yahoo.com}
         }}\\
    \vspace{0.8cm}

   {\it Department of Physics, Isfahan University of Technology \\
P.O.Box 84156-83111, Isfahan, IRAN, \\
School of Physics, Institute for Research in Fundamental Sciences (IPM)\\
P.O.Box 19395-5531, Tehran, IRAN.}

  \end{center}

  \vskip 2em

\begin{abstract}
We give the details of symplectic quantization for a system
containing second class constraints. This method is appropriate
for imposing infinite series of constraints due to the boundary
conditions. We use this method for massive bosonic strings in a
background B-field and find the correct expansions of the fields
in terms of the physical modes. We have found a canonical basis
for this model.
\end{abstract}

%\end{titlepage}

% \tableofcontents


%----------------------------------------------------------------------------------------------------------------
 \section{Introduction \label{sec1}}
Quantization process has been one of the most important problems
from the early days of quantum theory. According to the famous
prescription of Dirac\cite{Dirac}, given a classical theory with a
well behaved bracket in the phase space, the corresponding quantum
theory comes out as the result of changing the dynamical variables
to operators. Hence, the consistent algebraic structure of
classical brackets is an important part of the quantization
process. If the system possesses second class constraints one
needs to consider Dirac brackets instead of Poisson brackets.
However, this procedure should be consistent with the dynamics of
the system both in the classical and quantum levels. This demand
makes us to investigate the consistency of constraints in the
course of the time and finally find a consistent algebra of weakly
vanishing brackets among the constraints and the
Hamiltonian\cite{BatleGomis,Heno}.

On the other hand, we have difficulty with the Dirac method since
by restricting the system to live on the constraint surface one
may miss the beautiful Poisson structure of the phase space. In
other words, we need a new canonical structure that guarantees a
closed algebra of Poisson brackets on the constraint surface.
According to the famous "Darboux theorem" one can, in principal,
find a set of canonical coordinates in which the constraint
surface is described by a number of canonical conjugate pairs in
addition to some extra coordinates. Faddeev and Jackiw
\cite{Fad-Jack} used the above theorem, at each step of
consistency, to construct a new method for analyzing the
constrained systems. The Faddeev-Jackiw formulation and its
equivalence with the Dirac approach is studied in detail by
several authors, see for instance \cite{Shir-Moj,Castelani}.

A modified version of Faddeev-Jackiw formulation recognized as
"symplectic quantization" is more or less used by some authors,
when they want to build a quantum theory out of a given classical
action\cite{nelson,ruiz,Ho-Chu1,Ho-Chu2}.  As we will see, the
most essential point in this process is finding the complete set
of physical modes and the correct expansions of the fields in
terms of them. The main goal of this paper is giving the essential
aspects of symplectic quantization and applying it for
quantization of massive bosonic string in a background B-field as
a rich example. We will show that the method in its original form
works well for the bosonic string in background B-field with no
need to any modification such as the additional time averaging
suggested in \cite{Ho-Chu1}.

In this paper we first review the essential aspects of the
symplectic quantization in section 2. We restrict ourselves to a
purely second class system. This is specially the case when all
the constraints have originated from the boundary conditions. In
section 3 we consider in details the massive bosonic string again
in a background B-field. As we will see the constraint structure
of this case is more than a simple extension of the massless case.
Then we exhibit the power of the symplectic quantization method in
this interesting and illuminating example. The results on the
commutation relations would be discussed in the text and also in
the last section which is devoted to the concluding remarks.
\section{The Symplectic Approach \label{sec2}}
The Hamiltonian formulation of a dynamical system can be achieved
by using the first order Lagrangian
\begin{equation}
L_{F.O.}=\sum _i p_i\dot{q}_i-H(q,p).\label{a-0}
\end{equation}
If one uses general coordinates $\xi^i \ (i=1,...,2N)$ for the
phase space, which may be or may not be canonical, the first order
Lagrangian would be written as
\begin{equation}
L_{F.O.}=\sum _i a_i(\xi)\dot{\xi}^i-H(\xi).\label{a-2}
\end{equation}
In this form the action is
\begin{equation}
S=\int[a_i(\xi)d\xi^i-H(\xi)dt],\label{aa-3}
\end{equation}
where the kinetic term exhibit a one-form on the phase space
manifold. The Euler-Lagrange equations of motion can be written as
$\omega_{ij} \dot{\xi}^j = \partial H/\partial \xi^i$, where
$\omega_{ij}= \partial a_i/\partial \xi^j -\partial a_j/\partial
\xi^i$ is the symplectic matrix. It defines the "symplectic
two-form" as
\begin{equation}
\Omega=\frac{1}{2}\omega_{ij}d\xi^i\wedge d\xi^j=d a,\label{a-5}
\end{equation}
where $a$ is the kinetic one-form of the action (\ref{aa-3}). The
equations of motion can be solved as $ \dot{\xi}^i= \omega^{ij}
\partial H/\partial \xi^j$, where $\omega^{ij}$ is the inverse of
$\omega_{ij}$. These equations should be equivalent to the
canonical equations of motion $\dot{\xi}^i=\{\xi^i,H\}$, where
$\{\ ,\ \}$ means the Poisson bracket. Hence, for any two
functions $f(\xi)$ and $g(\xi)$ the Poisson bracket should be
defined as
\begin{equation}
\{f,g\}=\sum_{i,j=1}^N \frac{\partial f}{\partial\xi^i}
\omega^{ij}\frac{\partial g}{\partial\xi^j}\ .\label{a-8}
\end{equation}
As we see, the important role of the symplectic two-form is that
\emph{the inverse of symplectic matrix defines the Poisson
brackets in arbitrary coordinates of phase space}.

Now, suppose we are given a set of second class constraints
$\phi_a(\xi)\approx0$ on the phase space variables. Suppose this
set includes primary constraints as well as secondary constraints
which come out from the consistency of primary ones. In other
words, no new constraint may be added to the system as the result
of consistency of the present constraints. For instance, these
constraints can be considered as the infinite set of boundary
conditions and their consistency conditions, to be discussed in
the following section. Under imposing the constraints on arbitrary
coordinates $\xi^k$ of the original phase space, the reduced phase
space is described by the coordinates $\eta^a \quad a=1,..,2m \
(m<N)$, which are not necessarily canonical. On the reduced phase
space, we have $ \xi^k=\xi^k(\eta^a)$, and the induced symplectic
tensor is
\begin{equation}
\omega_{ab}=\frac{\partial \xi^i}{\partial \eta^a} \frac{\partial
\xi^j}{\partial \eta^b} \omega_{ij}.\label{a-12}
\end{equation}
The induced symplectic tensor $\omega _{ab}$ can be written by
imposing the constraints on the symplectic two-form $\Omega$.
Inverting $\omega _{ab}$, then gives the inverse tensor $\omega
^{ab}$ which determines the Dirac brackets on the reduced phase
space.

For a field theory  in $d+1$ dimensions with the dynamical fields
$\phi_s(x,t)$ the symplectic two-form in terms of the original
fields is given by
\begin{equation}
\Omega=\sum_s\int d^dx d\Pi_s(x,t)\wedge d\phi_s(x,t).
\label{a-14}
\end{equation}
where $\Pi(x,t)$ are momentum fields. Upon imposing a set of
second class constraints, such as initial boundary conditions and
their consistency conditions, suppose the phase space fields
$\phi_s(x,t)$ and $\Pi_s(x,t)$ can be written in terms of a
restricted set of physical mods $a_n(t)$. Inserting the
corresponding expansions of $\phi_s(x,t)$ and $\Pi_s(x,t)$ in Eq.
(\ref{a-14}), the symplectic two-form can be written, in
principal, as
\begin{equation}
\Omega=\sum_{n,m}\omega_{nm}da^n\wedge da^m. \label{a-15}
\end{equation}
The summations in (\ref{a-15}) are understood to include
integration over the continues variables in cases where physical
modes $a_n(t)$ depend on continues parameters. The inverse tensor
$\omega^{nm}$ defines the Dirac brackets of the reduced phase
space coordinates as
\begin{equation}
\{a^n,a^m\}=\omega^{nm} \ .\label{a-16}
\end{equation}
Note that the initial Poisson brackets is no longer valid on the
reduced phase space.

We emphasize the important point that up to this point the
symplectic quantization is done without solving the equations of
motion of the variables or fields. In other words, we have used
the first part of the action (\ref{aa-3}) so far. The dynamics is,
however, the responsibility of the second part, i.e. the
Hamiltonian. Properties of the Poisson structure on the classical
phase space, as well as the algebra of the operators in quantum
mechanics do not depend at all on the form of the Hamiltonian. For
example for a particle in one dimension the most important thing
for quantization is the classical bracket $\{x,p\}=1$ or
equivalently $\{a,a^\ast\}=\frac{1}{i\hbar}$; which converts to
$[x,p]=i\hbar$ or $[a,a^\dagger]=1$ upon quantization. These
relations do not depend on the explicit time dependence of a(t)
and $a^{\dagger}(t)$ to be for example $a(0)e^{i\omega t}$ and
$a^{\dagger}(0)e^{-i\omega t}$ for simple harmonic oscillator.

It worths reminding that upon ordinary conditions, the classical
brackets, as well as the quantum commutators, are independent of
time. To see this explicitly, suppose at time $t$ we have $\{
\xi^i(t),\xi^j(t)\}=\omega ^{ij}$. Assume $\omega ^{ij}$ is
independent of $\xi^{i_,}s$, which is the case in most of the
problems. In the course of the time, the variation of $\xi^i$ is
$\delta \xi^i=\{\xi^i,H\}\delta t$. Using the Jacobi identity the
bracket of $\xi^i$ and $\xi^j$ at time $t+\delta t$ is
\begin{eqnarray}
\left\{ \xi^i+\delta \xi^i,\xi^j+\delta \xi^j \right\}&=&\omega
^{ij}+\delta t\left[ \left\{ \{ \xi^i,H\},\xi^j \right\}+\left\{
\xi^j,\{\xi^i,H\}\right\} \right]\nonumber
\\ { } &=& \omega ^{ij}+\delta t\left\{ H,\{\xi^i,\xi^j\} \right\} \nonumber \\
&=& \omega^{ij}. \label{a-19}
\end{eqnarray}
\section{Massive bosonic string}
Consider the Lagrangian of the  massive bosonic string in an
external B-field,
\begin{equation}
L=\frac{1}{2}\int_0^l d\sigma \left[\dot{X}^2-X'^2-m^2
X^2+2B_{ij}\dot{X}_iX'_j\right] \ ,\label{a-34}
\end{equation}
where "dot" and "prime" represent differentiation with respect to
$\tau$ and $\sigma$ respectively and $X^2\equiv X_iX_i$, etc. This
is the simplified version of a model in which among the whole set
of bosonic fields $X^\mu$ an even number $X^i$ are influenced by a
background antisymmetric B-field and are attached at the
end-points to D-brains. Here, we have omitted those components of
$X^\mu$ which possess ordinary Neumann boundary conditions and are
not coupled to B-field. We consider the simplest case where
$i=1,2$, and the metric of the truncated target space is
Euclidian. Hence, we consider all the space indices as lower
indices. The antisymmetric B-field in two space dimensions can be
written as
\begin{equation}
B_{ij}\equiv \left( \begin{array}{cc} 0 & \tilde{B} \\
-\tilde{B} &0 \end{array} \right) . \label{b-3}
\end{equation}
The massless case ($m=0$) is studied in different aspects by
several authors \cite{Shir-Deh,Shir-Sheykh,witten,ardalan,Ho-Chu3}
with the well-known result of non commutativity at the end-points.
Demanding the variation of the action vanish under arbitrary
variation $\delta X_i$ gives the equation of motion as
$\left(\partial^2_{\tau}-
\partial^2_{\sigma}-m^2\right)X_i=0$, while the boundary
conditions are ${X_i}'+B_{ij}\dot{X}_j=0$ at $\sigma=0$ and
$\sigma=l$. The momentum fields are given by $P_i=\dot{X}_i+
B_{ij}X'_j$. The canonical Hamiltonian reads
\begin{equation}
H=\frac{1}{2}\int_0^l d\sigma \left[(P-BX')^2+X'^2+m^2X^2\right] \
.\label{a-36}
\end{equation} The fundamental Poisson brackets (before imposing the
constraints) read
\begin{eqnarray}&&\{X_i(\sigma,\tau),X_j(\sigma',\tau)\}=0,\nonumber\\
&&\{P_i(\sigma,\tau),P_j(\sigma,\tau)\}=0,\label{b-2}\\
&&\{X_i(\sigma,\tau),P_j(\sigma,\tau)\}=\delta_{ij}\delta
(\sigma-\sigma').\nonumber
\end{eqnarray}
Consider the boundary conditions as primary constraint
$\Phi_i^{(1)}\equiv \phi_i^{(1)}|_{\sigma=0}$ and
$\bar\Phi_i^{(1)}\equiv \phi_i^{(1)}|_{\sigma=l}$ where
\begin{equation}
\phi_i^{(1)}\equiv M_{ij}X'_j+B_{ij}P_j\ .\label{b-4}
\end{equation}
in which $M=1-B^2$. As indicated in details in references
\cite{Shir-Deh,Shir-Sheykh}, consistency of primary constraints in
time determines the Lagrange multipliers $\lambda_i$ and
$\bar\lambda_i$ in the total Hamiltonian $H_T=H_c
+\lambda_i\Phi_i^{(1)}+ \bar{\lambda}_i \bar{\Phi}_i^{(1)}$ to be
zero and at the same time gives second level constraints as
$\Phi_i^{(2)}\equiv \psi_i(0,\tau)$ and $\bar{\Phi}_i^{(2)}\equiv
\psi_i(l,\tau)$ where
\begin{equation}
\psi_i(\sigma,\tau)=\partial_{\sigma}P_i-m^2B_{ij}X_j.\label{a-41}
\end{equation}
Direct calculation shows that third and forth level constraints
are $(\partial^2_{\sigma}-m^2)\phi_i$ and
$(\partial^2_{\sigma}-m^2)\psi_i$ at the end-points, respectively.
In this way even and odd level constraints at $\sigma=0$ read
respectively as
\begin{eqnarray}
&&\Phi_i^{(2r)}=(\partial^2_{\sigma}-m^2)^{r-1}\psi_i\mid_{\sigma=0}\approx
0\;\ r\geq1 \ ,
\nonumber\\
&&\Phi_i^{(2r+1)}=(\partial^2_{\sigma}-m^2)^{r}\phi_i\mid_{\sigma=0}\approx
0\;\ r\geq0 \ ,\label{a-42}
\end{eqnarray}
with similar expressions for $\bar{\Phi}_i^{(2r)}$ and
$\bar{\Phi}_i^{(2r+1)}$ at $\sigma=l$.

Expanding the real fields $X(\sigma,\tau)$ and $P(\sigma,\tau)$ as
Fourier integrals, we have
\begin{eqnarray}&&X_i(\sigma,\tau)=\frac{1}{\sqrt{2\pi}}\int^{\infty}_{-\infty}dk
[a_i(k,\tau)\cos k\sigma+b_i(k,\tau)\sin k\sigma],\nonumber\\
&&P_i(\sigma,\tau)=\frac{1}{\sqrt{2\pi}}\int^{\infty}_{-\infty}dk
[c_i(k,\tau)\cos k\sigma+d_i(k,\tau)\sin k\sigma].\label{a-23}
\end{eqnarray}
It is clear that $a_i(k,\tau)$ and  $c_i(k,\tau)$ are even while
$b_i(k,\tau)$ and $d_i(k,\tau)$ are odd functions of $k$. Before
imposing the constraints all of the above modes are present. Using
the fundamental brackets (\ref{b-2}) one can find the following
Poisson brackets among the Fourier modes
\begin{eqnarray}
&&\{a_i(k,\tau),c_j(k',\tau)\}=\delta_{ij}\delta(k-k'),\nonumber \\
&&\{b_i(k,\tau),d_j(k',\tau)\}=\delta_{ij}\delta(k-k'),\label{a-46}
\end{eqnarray}
while all other bracket vanish. This shows that the canonical
pairs $\left(a_i(k,\tau),c_i(k,\tau)\right)$ and
$\left(b_i(k,\tau),d_i(k,\tau)\right)$ act as an alternative
canonical basis for the original phase space. Upon imposing the
constraints a large number of these modes will be omitted, leaving
a much smaller number of them as the canonical coordinates of the
reduced phase space.

We now impose the set of constraints (\ref{a-42}) on the Fourier
expansions (\ref{a-23}) to find
\begin{eqnarray}
&&\int_{-\infty}^{\infty} dk (-1)^r(k^2+m^2)^r\left[
B_{ij}c_j(k,\tau)+kM_{ij}b_j(k,\tau)\right]=0 \ ,\nonumber\\
&&\int_{-\infty}^{\infty} dk (-1)^r(k^2+m^2)^r\left[
kd_i(k,\tau)-m^2B_{ij} a_j(k,\tau)\right] =0 \ .\label{a-47}
\end{eqnarray}
Since the expressions in the brackets are even functions of $k$,
the conditions (\ref{a-47}) for arbitrary $r$ will be satisfied
only if $b(k,\tau)= -M^{-1}Bc(k,\tau)/k$ and $d(k,\tau)= m^{2}B
a(k,\tau)/k$. This leads to omitting the modes $b(k,\tau)$ and
$d(k,\tau)$ in favor of $a(k,\tau)$ and $c(k,\tau)$ respectively.
For simplicity we omit the indices $i,j,\cdots$ on the fields and
modes from now on. Imposing the constraints at the end point
$\sigma=l$ gives
\begin{eqnarray}
&&\int_{-\infty}^{\infty} dk (-1)^r(k^2+m^2)^r\left[\frac{1}{k}m^2
B^2 -kM\right] a(k,\tau) \sin kl=0 \ ,\nonumber\\
&&\int_{-\infty}^{\infty} dk
(-1)^r(k^2+m^2)^r\left[-k+\frac{1}{k}m^2 B^2M^{-1}\right]c(k,\tau)
\sin kl=0\ .\label{a-50}
\end{eqnarray}
These equations show that $a(k,\tau)$ and  $c(k,\tau)$ should
vanish except for $k=n\pi/l$ with integer $n$ or when
\begin{equation}
k^2=m^2 B^2 M^{-1}=-\frac{m^2 \tilde{B}^2}{1+\tilde{B}^2}\equiv
-k^2_0\ .\label{a-52}
\end{equation}
The first possibility leads to oscillatory modes as
\begin{eqnarray}
&&X_{\textrm{os}}=\sqrt{\frac{2}{l}}\sum_{n=1}^{\infty}
\left[a^{(n)}(\tau)\cos\frac{n\pi\sigma}{l}-\frac{l}{n\pi}M^{-1}
Bc^{(n)}(\tau)\sin\frac{n\pi\sigma}{l}\right]\ ,\nonumber \\
&&P_{\textrm{os}}=\sqrt{\frac{2}{l}}\sum_{n=1}^{\infty}\left[c^{(n)}(\tau)\cos
\frac{n\pi\sigma}{l}+\frac{l}{n\pi}m^2Ba^{(n)}(\tau)\sin
\frac{n\pi\sigma}{l}\right]\ .\label{a-53}
\end{eqnarray}
The normalization coefficients $\sqrt{2/l}$ are chosen for future
convenience. This choice also makes the correct dimensionality for
$a^{(n)}$ as $(\textrm{Mass})^{-1/2}$ and for $c^{(n)}$ as
$(\textrm{Mass})^{1/2}$. Note that the fields $X_i$ are
dimensionless.

The possibility (\ref{a-52}) corresponds to zero mode solutions
with $\sinh k_0 \sigma$ and $\cosh k_0 \sigma$. Traditionally the
zero mode solution is denoted as the zero frequency (infinite wave
length) limiting term in the Fourier expansions, as shown for
example for the massless case of the current problem in
\cite{Shir-Deh}. Here, however, we interpret \emph{the zero mode
solution as a solution which satisfies the boundary conditions not
only at the end-points but also throughout all the medium}. This
interpretation works well for constant term in the case of
ordinary Neumann boundary condition of free bosonic string as well
as for the zero mode terms for the massless string in B-field. To
see the details, suppose the first and second level constraints
are valid throughout all the string as the following coupled
deferential equations
\begin{eqnarray} &&BP+MX'=0\ ,\nonumber \\
&&P'-m^2BX=0 \ ,\label{a-55}
\end{eqnarray}
which gives $MX''+m^2B^2X=0\ $. Since $B^2=-\tilde{B}^2
\mathbf{1}$ and $M=(1+\tilde{B}^2)\mathbf{1}$ the most general
solutions of Eqs. (\ref{a-55}) can be chosen as
\begin{eqnarray}
&&X_{\textrm{zm}}(\sigma,\tau)=\frac{1}{\sqrt{l}}\left[a^{(0)}(\tau)\cosh
[k_0(\sigma -l/2)]-\frac{1}{k_0}M^{-1}Bc^{(0)}(\tau)\sinh
[k_0(\sigma -l/2)]
\right]\nonumber \\
&&P_{\textrm{zm}}(\sigma,\tau)=\frac{1}{\sqrt{l}}\left[c^{(0)}(\tau)\cosh
[k_0(\sigma -l/2)]+\frac{1}{k_0}m^{2}Ba^{(0)}(\tau)\sinh
[k_0(\sigma -l/2)]\right]\label{a-59}
\end{eqnarray}
where $k_0$ is given in Eq.(\ref{a-52}). Again we have imposed the
normalization coefficient $\sqrt{1/l}$ for making correct
dimensionality and future convenience. Note that we have chosen
the argument of sinh and cosh functions as measured from the
center of the string. Besides respecting the symmetry of the
Hamiltonian under $\sigma \rightarrow -\sigma$ when we shift the
origin of $\sigma$ to $l/2$, the reason for this choice is that
the physical modes in this sector turn out finally to be canonical
modes, as we will see in the near future. The most general
solution of the fields are superposition of zero-mode and
oscillatory solutions as
\begin{eqnarray}
&& X=X_{\textrm{zm}}+X_{\textrm{os}},\nonumber\\
&&P=P_{\textrm{zm}}+P_{\textrm{os}}.\label{a-60}
\end{eqnarray}

Now we want to quantize the theory using the symplectic approach.
After a long but direct calculation the symplectic two-form
$\Omega=\sum_i\int d\sigma dP_i(\sigma,\tau) \wedge
dX_i(\sigma,\tau)$ in terms of the physical modes $a^{(n)}(\tau)$
and $c^{(n)}(\tau)$  read
 \be \Omega = \frac{\sinh k_0l}{lk_0}
\left(dc_i^{(0)} \wedge da_i^{(0)}\right)
+\sum_{n=1}^{\infty}\left(1+\frac{l^2k_0^2 }
{n^2\pi^2}\right)\left(dc_i^{(n)}\wedge
da_i^{(n)}\right)\label{a-61}
 \ee
The important point in derivation of Eq.(\ref{a-61}) is that all
cross terms composed of zero-modes wedge oscillatory modes have
been canceled. This is a good news, since it makes the symplectic
matrix block diagonal with one $4\times 4$ block for the zero
modes and an infinite dimensional block for the oscillatory modes.
Moreover in oscillatory modes there is no mixing between differen
oscillators (i.e. different $n$). In zero mode sector also we do
not encounter cross terms such as $da_i^{(0)} \wedge da_j^{(0)}$
and $dc_i^{(0)}\wedge dc_j^{(0)}$. This is because of our suitable
choice of zero modes in Eqs (\ref{a-59}) as combinations of $\sinh
[k_0(\sigma -l/2)]$ and $\cosh [k_0(\sigma -l/2)]$. These
opportunities make it easy to find the inverse of symplectic
matrix and write down the brackets of physical modes. Hence, the
non-vanishing Dirac brackets among physical modes are as follows
 \be
[a_i^{(n)} , c_j^{(s)}]  =  N_n^{-1}\delta_{ij}\delta_{ns},
\label{a-62}
 \ee
 where
 \be
N_0 \equiv \frac{\sinh (k_0l)}{k_0l} ,\hspace{1cm} N_n \equiv
1+\frac{k_0^2l^2}{n^2\pi^2}
 \ \ n\ne 0\ . \label{d-10}
 \ee
 Note that we did not need to solve the equations of motion
up to this point. The theory can be quantized, without any need to
full solutions of the equations of motion by converting the
brackets (\ref{a-62}) into the quantum commutators. Similar to the
massless case, one can find the time dependence of the physical
modes by directly solving their equations of motion. For this aim
it is more economical to write the canonical Hamiltonian
(\ref{a-36}) in terms of the physical modes. The result is
 \be
H= \frac{1}{2} \sum_i\sum_{n=0}^\infty N_n\left[
M^{-1}[c^{(n)}_i]^2+ M\omega_n^2[a^{(n)}_i]^2 \right]\ ,
\label{c-3}
 \ee
where
 \be
\omega_0^2 \equiv m^2 M \ , \hspace{1cm} \omega_n^2 \equiv
m^2+\frac{n^2\pi^2}{l^2} \ \ n\ne 0 \label{c-4}
 \ee
The Hamiltonian (\ref{c-3}) is a superposition of infinite number
of independent harmonic oscillators. The canonical equations of
motion can be solved for $n=0,1,\cdots$ as
\begin{eqnarray}
&& a^{(n)}(\tau)=a^{(n)}(0)\cos \omega_n \tau +
\frac{1}{M\omega_n}c^{(n)}(0)\sin \omega_n \tau\ , \nonumber \\
&& c^{(n)}(\tau) = c^{(n)}(0)\cos \omega_n \tau - M\omega_n
a^{(n)}(0) \sin \omega_n \tau \ ,\label{c-5}
\end{eqnarray}
where $a^{(n)}(0)$ and $c^{(n)}(0)$ are Schrodinger modes.
Inserting $a^{(n)}(\tau)$ and $c^{(n)}(\tau)$ from Eqs.
(\ref{c-5}) in the expansions (\ref{a-53}) and (\ref{a-59}) of the
fields, gives in fact the solutions of the equations of motion
consistent with the boundary conditions, written in terms of the
canonical Schrodinger modes.

Our results here are different from those of reference
\cite{Ho-Chu2} in the following aspects:

1) The algebra of physical modes $a^{(n)}(\tau)$ and
$c^{(n)}(\tau)$ is much simpler. They are canonical conjugate
pairs with position-momentum like brackets given in (\ref{a-62}).

2) Inserting our final expansions of the fields in terms of the
Schrodinger modes in the symplectic two-form leads again to
expression (\ref{a-61}) for the initial values of the modes. In
other words, there is no explicit time dependence as expected.
However, direct calculation shows that, similar to the massless
case of reference \cite{Ho-Chu1}, the symplectic two-form
constructed from expansions given in \cite{Ho-Chu2} contains time
dependent terms.

It seems that real functions for spatial dependence of the fields
should be accompanied naturally with real time dependent of the
modes as seen in (\ref{c-5}). Combining spatial real functions
with the time dependence of the form $e^{i\omega \tau}$ may lead
to unwanted time dependence in the symplectic two-form. In fact,
this time dependence is the origin of an unnecessary step of time
averaging of the symplectic two-form suggested in reference
\cite{Ho-Chu1}.

In order to complete our discussions and enlighten some unclear
points of the literature let us calculate the Dirac brackets of
the original fields to see the effect of the B-field and boundary
conditions on the commutativity of fields. To do this using the
brackets (\ref{a-62}) and Eqs.(\ref{a-53}) ,(\ref{a-59}) and
(\ref{a-60}) we can compute the equal time Dirac brackets of the
original coordinates and momentum fields as follows
\begin{eqnarray}
&&[X_i(\sigma,\tau),X_j(\sigma',\tau)]=2(BM^{-1})_{ij}f(\sigma+\sigma')
\ , \label{b-9} \label{b-12}\\
&&[P_i(\sigma,\tau),P_j(\sigma',\tau )] =
2m^2B_{ij}f(\sigma+\sigma') \ ,\label{b-13}\\
&&[X_i(\sigma,\tau),P_j(\sigma',\tau)]
=\delta_{ij}g(\sigma,\sigma') \label{a-63}
\end{eqnarray}
where
\begin{equation}
f(\sigma)=\frac{\sinh [k_0(\sigma-l)]}{\sinh k_0l}
+\frac{2}{\pi}\sum_{n=1}^{\infty}\frac{1}{n}\left(1+\frac{l^2k_0^2}
{n^2\pi^2}\right)^{-1}\sin \left( \frac{n\pi\sigma}{l} \right)\
,\label{a-64}
\end{equation}
and
\begin{eqnarray}
g(\sigma,\sigma') &=&
2k_0\frac{\cosh [k_0(\sigma+\sigma'-l)]}{\sinh k_0l}\nonumber\\
{ }& { } &+\frac{4}{l}\sum_{n=1}^{\infty}\left(1+\frac{l^2k_0^2}
{n^2\pi^2}\right)^{-1} \cos \frac{n\pi\sigma}{l} \cos
\frac{n\pi\sigma'}{l}\nonumber \\
{ } & { } & +\frac{4}{l}\sum_{n=1}^{\infty}\frac{l^2k_0^2}
{n^2\pi^2}\left(1+\frac{l^2k_0^2} {n^2\pi^2}\right)^{-1}\sin
\frac{n\pi\sigma}{l}\sin \frac{n\pi\sigma'}{l}.\label{b-14}
\end{eqnarray}
Let us first consider the brackets (\ref{b-12}) and (\ref{b-13})
in details. The function $f(\sigma)$ reduces, for $m=0$, to
\begin{equation}
f_0(\sigma)=-1+\frac{\sigma}{l}+\frac{2}{\pi}\sum_{n=1}^{\infty}\frac{1}{n}
\sin \left( \frac{n\pi\sigma}{l} \right)\ ,\label{b-7}
\end{equation}
which gives
\begin{eqnarray}
&&\{X_i(\sigma,\tau),X_j(\sigma',\tau)\}=0 \;\;\
\sigma,\sigma'\neq0 , \nonumber \\
&& \{X_i(0,\tau),X_j(0,\tau)\}=-2(M^{-1}B)_{ij} ,\label{a-29}\\
&& \{X_i(l,\tau),X_j(l,\tau)\}=2(M^{-1}B)_{ij} . \nonumber
\end{eqnarray}
which is the standard results of non commutativity of the
end-points of a massless string in the background B-field. For $m
\ne 0$ using the Fourier expansion
\begin{equation}
\sinh [k_0(\sigma-l)]=\frac{2\sinh k_0l}{\pi}\sum_{n=1}^{\infty}
\frac{1}{n}\left( \frac{n^2\pi^2}{n^2\pi^2+k_0^2l^2}\right)\sin
\left( \frac{n\pi\sigma}{l} \right)\ , \label{b-10}
\end{equation}
in the interval $[0,2l]$, it is easily seen that $f(\sigma)$
vanishes for every value of $\sigma$ in the open interval $
(0,2l)$. For $\sigma=0$ and $\sigma=2l$, however, the Fourier
expansion (\ref{b-10}) is no longer valid since, beyond the
interval $(0,2l)$, the hyperbolic function on the left hand side
of Eq. (\ref{b-10}) can not be expanded in terms of periodic
functions $\sin \left( n\pi\sigma/l \right)$. Similarly, the last
expression in the right hand side of Eq. (\ref{b-7}) is the
expansion of the non-continues sawtooth function with
non-continuity on the points $\sigma=2kl$ for integer $k$. At the
end-points $\sigma=0$ and $\sigma=2l$ the function $f_0(\sigma)$
is +1 and -1 respectively from its definition in Eq. (\ref{a-64}).
The above argument can be used exactly in the same way for
$f(\sigma)$ given in Eq. (\ref{a-64}). In other words, $f(\sigma)$
in Eq. (\ref{a-64}), for arbitrary $k_0$, is the same as given in
Eq. (\ref{b-7}) for the case $k_0=0$. Hence we have
\begin{equation}
f(\sigma)=0 \ \ \textrm{for} \ \ 0<\sigma<2l, \quad f(0)=-1, \quad
f(2l)=1\ .\label{b-11}
\end{equation}
This result shows that non-commutativity of the coordinate fields
at the end points are exactly the same for massless and massive
cases.

As we see from Eqs. (\ref{b-9}) and (\ref{b-13}), for $m\ne 0$,
the momentum fields as well as the coordinate fields are
noncommutative at the end-points. This is due to dependence of
momentum fields to coordinate Fourier modes $a_i^{(n)}$ in
contrast to massless case which is just composed of momentum
Fourier modes $c_i^{(n)}$. It worth remembering that mixing of
coordinate and momentum Fourier modes in the expansions of
$X(\sigma,\tau)$ and $P(\sigma,\tau)$ is basically due to mixed
boundary conditions which is in turn resulted from the presence of
B-field.
%As is well-known, in the standard noncommutative models
%the main interest is when coordinates are non commutative and the
%momenta are remained commutative. Therefore it seems that the
%massive bosonic field can not provide an attractive model for the
%standard non-commutativity.

Next we consider the bracket of a coordinate and a momentum field
in Eq. (\ref{a-63}). The function $g(\sigma,\sigma')$ given in Eq.
(\ref{b-14}) can be written as
\begin{equation}
g(\sigma,\sigma')=\frac{4}{l}\sum_{n=1}^{\infty}\cos
\frac{n\pi\sigma}{l} \cos \frac{n\pi\sigma'}{l}+
\bar{g}(\sigma+\sigma')\ ,\label{b-15}
\end{equation}
where
\begin{equation}
\bar{g}(\sigma)= 2k_0\frac{\cosh [k_0(\sigma-l)]}{\sinh
k_0l}-\frac{4}{l}\sum_{n=1}^{\infty}\left(\frac{l^2k_0^2}
{n^2\pi^2+l^2k_0^2}\right) \cos \frac{n\pi\sigma}{l} \
.\label{b-16}
\end{equation}
Using the Fourier expansion
\begin{equation}
\cosh [k_0(\sigma-l)]=\frac{\sinh k_0l}{k_0l}+\sum_{n=1}^{\infty}
\left( \frac{2k_0l\sinh k_0l}{n^2\pi^2+k_0^2l^2}\right) \cos
\left( \frac{n\pi\sigma}{l} \right) \ , \label{b-17}
\end{equation}
for $\sigma \epsilon [0,2l]$, it is easily seen that
$\bar{g}(\sigma)=2/l$. So we have finally
\begin{equation}
g(\sigma,\sigma')=\frac{2}{l}+\frac{4}{l}\sum_{n=1}^{\infty}\cos
\frac{n\pi\sigma}{l} \cos
\frac{n\pi\sigma'}{l}=\delta(\sigma-\sigma')\ .\label{b-18}
\end{equation}
The interesting point is that the result is again independent of
$m$ or $k_0$. In other words, the fields $X(\sigma,\tau)$ and
$P(\sigma,\tau)$ are still canonical conjugate pairs in the
reduced phase space, for the massive case.

\section{Concluding Remarks}
Our main goal in this paper is to reintroduce "the symplectic
quantization method", in such a way that is appropriate for
imposing the second class constraints originated from the boundary
conditions. We showed that the essential point is imposing the set
of constraints on some "appropriate expansion" of the fields in
terms of suitable modes. Although the full dynamics of the system
is not required in order to use the machinery of the symplectic
approach, it is necessary to investigate the dynamics of the
constraints; which is, in fact, the program of finding the
constraint structure of the system.

The main outcome of this procedure is recognizing all independent
physical degrees of freedom, i.e. the coordinates of the reduced
phase space, and finding expansions of quantities of interest,
such as the original fields and the Hamiltonian in terms of them.
Note that in this viewpoint the mode expansions of the fields are
not just combinations of the solutions of the equations of motion
with some meaningless coefficients. These coefficients, if
precisely chosen, should coincide with the initial values of the
physical modes (i.e. the Schrodinger modes). Clearly, we would be
more happy if these coordinates constitute a canonical basis for
the reduced phase space.

%We considered massless and massive bosonic strings in the
%background B-field as two main examples to show how the symplectic
%method survives a systematic approach for quantization.

%After a brief review of the constraint structure of the ordinary
%massless bosonic string in B-field, we showed that the original
%program of the symplectic approach without any modification
%survives for the purpose of quantization, provided that the modes
%of the reduced phase space have been chosen carefully. Our mode
%expansions for the fields differ from those of reference
%\cite{Ho-Chu1}. However, our results  about non-commutativity of
%the coordinate fields at the end-points of the string are in
%agreement the well-known results of the literature such as
%\cite{Ho-Chu1,Shir-Deh,Shir-Sheykh,witten,ardalan}.

We then considered the model of a massive string in a background
B-field as an interesting and nontrivial example for applying the
idea of considering boundary conditions as Dirac constraints as
well as using the symplectic approach for quantizing the model.
Besides characteristics such as arising mixed boundary conditions
(Eqs. \ref{b-4} and \ref{a-41}) and the appearance of two sets of
infinite number of constraints at each boundary (Eq. \ref{a-42}),
which is common with the massless case, the problem of massive
bosonic string has its special attractions for two following
reasons;

i) The boundary conditions (\ref{b-4}) and (\ref{a-41})
incorporate coordinate and momentum fields almost on the same
footing. This results to expansions (\ref{a-53}) and (\ref{a-59}),
in which canonical conjugate pairs are present both in
$X(\sigma,\tau)$ and $P(\sigma,\tau)$, which finally leads to
noncommutative momentum fields (see Eq. \ref{b-13}) as opposed to
the massless case.

ii) Upon imposing the constraints, the massive bosonic fields
acquire nontrivial zero modes (see Eqs. \ref{a-59}) which can not
be derived from the limiting case of oscillatory modes. We suggest
to define the zero mode as solution of generalization of the
boundary conditions to the whole medium, instead of the boundaries
alone. Such a solution clearly satisfies the required conditions
at the boundaries and should be included in the most possible
expansions of the fields. We showed that this inclusion could be
happen naturally and need not to be added by hand (see our
discussion after Eq. \ref{a-50}).

Our main result for this model is finding a canonical basis for
the reduced phase space with a canonical algebra given in
(\ref{a-62}), which apart from being physically meaningful, is
much simpler to work with, compared to that of reference
\cite{Ho-Chu2}. We showed also that the Hamiltonian of the system
is simply the superposition of harmonic oscillators constructed
over these coordinate-momentum pairs.

Giving all technical details, we showed that the coordinate and
momentum fields are non-commutative at the end-points (Eqs.
\ref{b-12}-\ref{b-13} and \ref{b-11}); while they remain, after
imposing the constraints, canonically conjugate to each other in
the bulk of string (Eqs. \ref{a-63} and \ref{b-18}).

\textbf{Acknowledgment}

The authors thank M. M. Sheikh-Jabbari and M. Dehghani for their
valuable comments.

%%%%%%%%%%%%%%%%%%%%%%%%%%%%%%%%%%%%%%%%%%%%%%%%%%%%%%%%%%%%%%%%%%%%%%%
\end{document}